\documentclass[notitlepage,twocolumn,showpacs,preprintnumbers,amsmath,amssymb,superscriptaddress, prl,floatfix]{revtex4-1}
\usepackage{bm}
\usepackage{graphicx}
\usepackage{dcolumn}
\usepackage{tabularx}
\usepackage{bm}
\usepackage{epstopdf}
\usepackage{amsmath}
\usepackage{color}
\usepackage{here}
\begin{document}

\preprint{Preprint}

\title{Low- and high-$\beta$ lasers in class-A limit: photon statistics, linewidth, and the laser-phase transition analogy}

\author{N. Takemura}
\email[E-mail: ]{naotomo.takemura.ws@hco.ntt.co.jp}
\affiliation{Nanophotonics Center, NTT Corporation, 3-1 Morinosato-Wakamiya, Atsugi, Kanagawa 243-0198, Japan}
\affiliation{NTT Basic Research Laboratories, NTT Corporation, 3-1 Morinosato-Wakamiya, Atsugi, Kanagawa 243-0198, Japan}
\author{M. Takiguchi}
\affiliation{Nanophotonics Center, NTT Corporation, 3-1 Morinosato-Wakamiya, Atsugi, Kanagawa 243-0198, Japan}
\affiliation{NTT Basic Research Laboratories, NTT Corporation, 3-1 Morinosato-Wakamiya, Atsugi, Kanagawa 243-0198, Japan}
\author{M. Notomi}
\affiliation{Nanophotonics Center, NTT Corporation, 3-1 Morinosato-Wakamiya, Atsugi, Kanagawa 243-0198, Japan}
\affiliation{NTT Basic Research Laboratories, NTT Corporation, 3-1 Morinosato-Wakamiya, Atsugi, Kanagawa 243-0198, Japan}

\begin{abstract}
Nanocavity lasers are commonly characterized by the spontaneous coupling coefficient $\beta$ that represents the fraction of photons emitted into the lasing mode. While $\beta$ is conventionally discussed in semiconductor lasers where the photon lifetime is much shorter than the carrier lifetime (class-B lasers), little is known about $\beta$ in atomic lasers where the photon lifetime is much longer than the other lifetimes and only the photon degree of freedom exists (class-A lasers). We investigate the impact of the spontaneous coupling coefficient $\beta$ on lasing properties in the class-A limit by extending the well-known Scully-Lamb master equation. We demonstrate that, in the class-A limit, all the photon statistics are uniquely characterized by $\beta$ and that the laser phase transition-like analogy becomes transparent. In fact, $\beta$ perfectly represents the ``system size" in phase transition. Finally, we investigate the laser-phase transition analogy from the standpoint of a quantum dissipative system. Calculating a Liouvillian gap, we clarify the relation between $\beta$ and the continuous phase symmetry breaking.
\end{abstract}

\maketitle

\section{Introduction}
Nanocavity lasers in a cavity-QED regime are commonly characterized by the spontaneous emission coupling coefficient $\beta$, which represents the fraction of photons spontaneously emitted into a lasing mode. Recent technological progress has pushed $\beta$ close to unity \cite{Jin1994,Ulrich2007,Elvira2011,Ota2014,Takiguchi2016,Jagsch2018}, where the pump-input and light-output curves are  almost linear. Furthermore, the impact of $\beta$ on photon statistical properties is drawing considerable theoretical attention \cite{Rice1994,Druten2000,Gies2007,Chow2014,Mork2018,Vyshnevyy2018,Lohof2018}. However, since there are various parameters in lasers, it is difficult to extract the pure effect of $\beta$ on photon statistics. For example, there are active debates on whether or not high-$\beta$ lasers are helpful for Poissonian light emission with small pump power \cite{Elvira2011,Mork2018,Lohof2018}. In this paper, we investigate $\beta$ in the limit where only the photon degree of freedom exists, which is called the class-A regime. This limit is a rather ideal limit, but the meaning of $\beta$ is clarified in it. 

The coefficient $\beta$ is commonly used for class-B lasers, typically semiconductor lasers, where the photon lifetime is comparable to or smaller than the population inversion (carrier) lifetime. Thus, the dynamics are described by two degrees of freedom, namely photon and population inversion (carrier). On the other hand, $\beta$ is rarely discussed in relation to class-A lasers. Although a few pioneering papers have discussed $\beta$ in class-A lasers \cite{Jin1994,Yamamoto1983}, several fundamental questions remain. How is $\beta$ defined in class-A lasers? How does $\beta$ affect second-order correlation function $g^{(2)}(\tau)$ and laser linewidth $g^{(1)}(\tau)$? How, is $\beta$ interpreted in the quantum laser-phase transition analogy? To answer these questions, we employ a simple but very rich model: the Scully-Lamb master equation.

First, we study the class-A limit of conventional class-B rate equations and obtain a class-A rate equation, which leads to the Scully-Lamb birth-death master equation. With this approach, we establish a connection between class-A and -B lasers and clarify the meaning of $\beta$ in class-A lasers. Using the master equation, we find that the spontaneous emission coupling coefficient $\beta$ uniquely characterizes both static and dynamic photon statistics. Furthermore, we find that the relation between $\beta$ and the effective "system size" \cite{Rice1994} becomes evident in the class-A limit, which is because the detailed balance condition is satisfied in this limit. For instance, thresholdless lasers with $\beta=1$ do not exhibit any phase transition-like signature in terms of photon statistics, which is understood as a breakdown of the phase transition in a ``small system". Second, we study laser linewidth using the original Scully-Lamb master equation  \cite{Scully1967}, which includes the off-diagonal part of the density matrix. In contrast to the case with photon statistics, linewidth narrowing will occur even for thresholdless class-A lasers with $\beta=1$.

Finally, we discuss the lasing phenomena from the standpoint of the quantum dissipative phase transition. For this purpose, the Liouvillian gap, which is defined as a nonzero eigenvalue of a Liouvillian closest to zero, plays a key role. When an open dissipative phase transition occurs, the Liouvillian gap is expected to close. Our finding is that the Liouvillian gap of the Scully-Lamb master equation  corresponds to the linewidth. Furthermore, we argue that it may be possible to understand the Liouvillian gap of the Scully-Lamb master equation  in the framework of the dissipative second-order phase transition with symmetry recently studied in \cite{Minganti2018}. In our case, the Scully-Lamb master equation has a continuous phase symmetry.

\section{Rate equation}
\subsection{Classification of lasers}
Laser systems where atoms (carriers) are interacting with photons confined in a cavity are typically characterized by the light-matter interaction strength $g$, dephasing rate $\gamma_\perp$, population decay rate $\gamma_\|$, and photon decay rate from the cavity $\gamma_c$. Arrechi proposed the following classification of lasers based on the time scales of the three different decay rates \cite{Arecchi1984,Arecchi2012}: $\gamma_\perp$, $\gamma_A$, and $\gamma_c$.

Class-A ($\gamma_\perp,\gamma_A\gg\gamma_c$): The atom (carrier) degree of freedom is adiabatically eliminated and the dynamics is governed only by the photon degree of freedom. In this paper, we consider this limit. Gas and dye lasers are usually classified into this category.

Class-B ($\gamma_\perp\gg\gamma_c\gtrsim\gamma_\|$): In this case, only the polarization degree of freedom is adiabatically eliminated, and the dynamics are described with the photon and atom (carrier) population. The simplest model for the class-B lasers is given by the Statz-deMars rate equations. Since the spontaneous coupling coefficient $\beta$ is conventionally defined in this framework, we start from the Statz-deMars rate equations and consider the class-A limit. Note that most semiconductor and solid-state lasers are class-B lasers.

\subsection{Rate equations}
\begin{figure}
\centering
\includegraphics[width=0.45\textwidth]{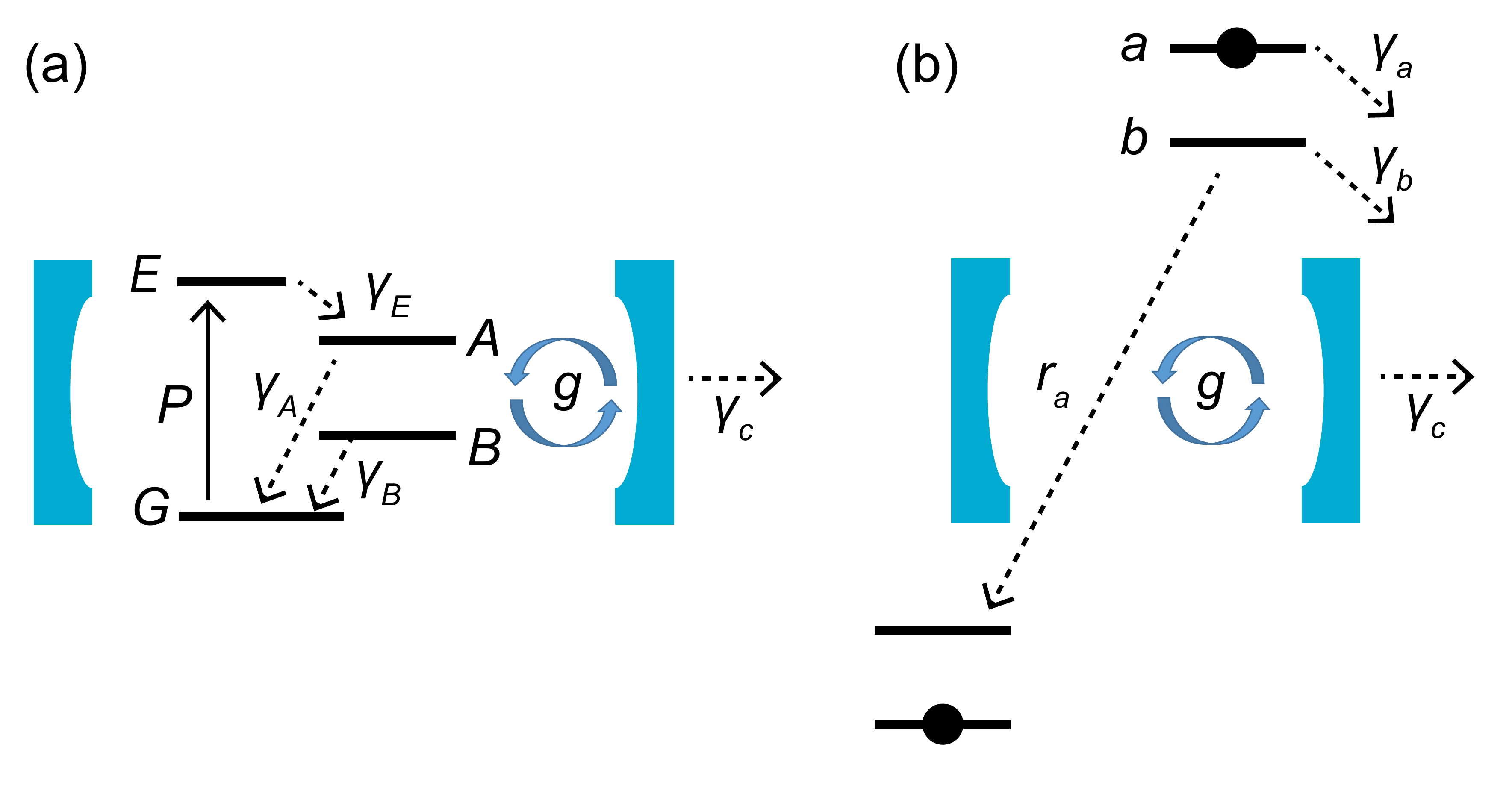}
\caption{(a) Schematic of rate equations Eq. (\ref{eq:REp}) and (\ref{eq:REn}), which are an ensemble of four-level atoms inside a cavity. The transition between the upper $A$ and lower level $B$ is coupled to photons in the cavity with a coupling strength $g$. $P$ represents pumping rate. The decay rate of levels $E$ and $B$ are very large; thus, the population of level $B$ is immediately depleted and repumped to level $A$. (b) Schematic of the Scully-Lamb quantum theory of lasers. Two-level atoms with upper $a$ and lower level $b$ are injected into a cavity at a rate $r_a$. The decay rates $\gamma_a$ and $\gamma_b$ represent the damping times of the upper ($a$) and lower ($b$) levels, respectively. A two-level atom interacts with photons with an interaction strength $g$.}
\label{fig:rate_eq}
\end{figure}

We consider an ideal four-level laser as illustrated in Fig. \ref{fig:rate_eq} \cite{Druten2000,Arkhipov2019}. Only the transition between the $A$ and $B$ levels is coupled to the cavity. We refer to levels $A$ and $B$ as upper and lower levels, respectively. Here, we assume that the depletion rate of the lower level population $\gamma_B$ is very large; thus, the population of the lower level is negligible and the population inversion $N$ is approximately the same as the upper level population. Moreover, since the decay rate $\gamma_E$ is also assumed to be sufficiently large, the population pumped at level $E$ is immediately transferred to the upper level $A$. With this assumption, the time evolutions of the photons $n$ and the population inversion $N$ may follow the conventional Statz-deMars rate equations for semiconductor lasers \cite{Bjork1991,Rice1994,Druten2000,Elvira2011}:
\begin{eqnarray}
\dot{n}&=&-\gamma_cn+\beta\gamma_\|N(n+1)\label{eq:REp}\\
\dot{N}&=&-\gamma_\|N-\beta\gamma_\|Nn+P\label{eq:REn},
\end{eqnarray}
where $P$ is the pumping rate, and $\gamma_c$ and $\gamma_\|$ represent the photon and population inversion decay rate, respectively. The coefficient $\beta$ is called the spontaneous emission coupling coefficient. These rate equations are frequently used for describing the dynamics of semiconductor lasers. It is important to note that the population inversion decay rate $\gamma_\|$ and the spontaneous coupling coefficient $\beta$ are given by \cite{Rice1994}
\begin{equation}
\gamma_\|\equiv\gamma_A+{4g^2}/{\gamma_\perp}\label{eq:gamma2}
\end{equation}
and
\begin{equation}
\beta\equiv\frac{4g^2}{\gamma_\|\gamma_\perp}=\frac{{4g^2}/{\gamma_\perp}}{\gamma_A+{4g^2}/{\gamma_\perp}}\label{eq:beta},
\end{equation}
where $\gamma_A$ represents the decay rates of the upper level $A$, while $\gamma_\perp$ is the dephasing rate of the dipole between the $A$ and $B$ states. 
Note that $\gamma_A$ includes all decay processes of the upper state $A$ that do not emit photons into the cavity mode (e.g. non-radiative decay).
Here, for simplicity, we assume that the cavity frequency is resonant to the dipole transition (zero detuning). The rate equations, Eq. (\ref{eq:REp}) and (\ref{eq:REn}), are derived from the Maxwell-Bloch equations by adiabatically eliminating the polarization degree of freedom, which is based on the fast dephasing rate: $\gamma_\perp\gg\gamma_c,\gamma_A$. Another important point is that the spontaneous emission effect is phenomenologically included in the rate equations, which is achieved by replacing $Nn$ with $N(n+1)$ in the derivation. 

From Eq. (\ref{eq:beta}), the spontaneous emission coupling coefficient $\beta$ is viewed as a fraction of the carrier decay into the lasing mode. Equation (\ref{eq:beta}) also indicates that $\beta=1$ is achieved when $\gamma_A=0$, where all the carriers decay into the lasing modes. Furthermore, the population inversion decay rate $\gamma_\parallel$ [see Eq. (\ref{eq:gamma2})] is enhanced by the dipole-photon coupling $g$ \cite{Mork2018}. Actually, the enhancement factor $4g^2/\gamma_\perp$ is the large dephasing limit ($\gamma_\perp\gg\gamma_c/2$) of the generalized Purcell enhancement factor: $4g^2/(\gamma_c/2+\gamma_\perp)$  \cite{Auffeves2009,Auffeves2010}. It is known that when the dephasing rate $\gamma_\perp$ is very large, the Purcell enhancement does not depend on the cavity decay rate (Q-factor) \cite{Sumikura2016}. We also note that even when $\gamma_A=0$ ($\beta=1$), the population inversion decays only with this Purcell effect. The enhancement of the rate $\gamma_\parallel$ plays an important role when we consider required conditions for high-$\beta$ class-A lasers (see Section 2D)

From the rate equations Eqs. (\ref{eq:REp}) and (\ref{eq:REn}),  the steady state photon number $\bar{n}$ is simply obtained as
\begin{eqnarray}
\bar{n}&=&\frac{1}{2{\beta}}\left[ -(1-{\beta}\tilde{P})+\sqrt{(1-{\beta}\tilde{P}) ^2+4{\beta}^2\tilde{P}}\ \right],\label{eq:nsteady}
\end{eqnarray}
where $\tilde{P}$ is the pump power normalized by the photon decay rate:
\begin{equation}
\tilde{P}=P/\gamma_c.\label{eq:normpump}
\end{equation}
Equation (\ref{eq:nsteady}) represents the pump-input and light-output curves  and indicates that the curve  has a kink at $P=P_{\rm th}$, where the kink threshold $P_{\rm th}$ is defined as
\begin{equation}
P_{\rm th}=\gamma_c/\beta,
\end{equation}
at which pump power the number of photons inside the cavity becomes $\beta^{-1/2}$. When $\beta=1$, the pump-input and light-output curve  [Eq. (\ref{eq:nsteady})] becomes linear as $\bar{n}=P/\gamma_c$ and a laser with $\beta=1$ is called ``thresholdless". Since we are neglecting the lower level population, the kink threshold $P_{\rm th}$ coincides with the widely used definition of lasing threshold, namely the pump power at which the gain becomes equal to the cavity loss \cite{Takemura2019}. Furthermore, the shape of the pump-input and light-output curves  is completely determined by $\beta$.

\subsection{Class-A rate equation}
Let us consider the class-A limit where the photon lifetime is much longer than the carrier lifetime: $\gamma_\|\gg\gamma_c$. In this limit, the adiabatic elimination of the carrier population reduces the two rate equations to a single equation of photons. Setting $\dot{N}=0$ in Eq. (\ref{eq:REn}), the carrier number adiabatically follows the photon number as
\begin{eqnarray}
N&=&\frac{P/\gamma_\|}{\beta n+1}.\label{eq:Nsteady}
\end{eqnarray}
Substituting Eq. (\ref{eq:Nsteady}) into Eq. (\ref{eq:REp}), we obtain the photon rate equation
\begin{equation}
\dot{n}=-\gamma_cn+\frac{\beta(n+1)}{\beta n+1}P,\label{eq:re_classA}
\end{equation}
This photon rate equation is very similar to the semi-classical rate equation derived by Lamb \cite{Scully1999,Loudon1973} but includes the spontaneous emission effect, which is represented by the $n+1$ term in the numerator of the second term on the right-hand side (RHS) of Eq. (\ref{eq:re_classA}). Thus, a photon buildup can occur even without a photon in the initial state.

\subsection{Conditions for high-$\beta$ class-A lasers}
Before introducing the master equation, we discuss the requirements for realizing class-A lasers that can be expressed with the class-A rate equation, Eq. (\ref{eq:re_classA}). In particular, whether or not high-$\beta$ and class-A conditions can be satisfied simultaneously is not trivial. First, the adiabatic elimination of the polarization degree of freedom from the Maxwell-Bloch equations requires
\begin{equation}
\gamma_\perp\gg\gamma_A\ \ {\rm and}\ \ \gamma_\perp\gg\gamma_c,\label{condition0}
\end{equation}
which is necessary for obtaining the rate equations, Eqs. (\ref{eq:REp}) and  (\ref{eq:REn}). Second, the photon lifetime must be much longer than the population lifetime
\begin{equation}
\gamma_\|\gg\gamma_c\ \ {\rm (class\mathchar`-A}\ {\rm limit)}.\label{condition1}
\end{equation}
Finally, since lasers are operating in the weak-coupling regime, we impose a weak-coupling condition by assuming that the dephasing rate is larger than the coupling between the polarization and photons
\begin{equation}
\gamma_\perp>g\ \ {\rm (weak}\ {\rm coupling)}.\label{condition2}
\end{equation}
Now, let us recall that the decay rate $\gamma_\|$ is enhanced by the Purcell effect indicated as Eq. (\ref{eq:gamma2}). For $\beta=1$, thus $\gamma_A=0$, the carrier decay rate is given by
\begin{equation}
\gamma_\|=\frac{4g^2}{\gamma_\perp}\label{condition3}
\end{equation}
Combining Eq. (\ref{condition0})-(\ref{condition3}), thresholdless class-A lasers ($\beta=1$) must satisfy the condition
\begin{equation}
\gamma_\perp^2>g^2\gg\gamma_c\gamma_\perp\ \ {\rm for}\ \beta=1.\label{condition_rate}
\end{equation}
With this condition, we can realize high-$\beta$ class-A lasers. 
To achieve $\gamma_A\simeq0$, first, the non-radiative decay must be minimized. Second, $\gamma_A$ can be reduced by enhancing the spontaneous emission rate of the upper level a into the cavity mode with the Purcell effect. Recalling that the Purcell factor is inversely proportional to the mode-volume of the cavity \cite{Anonymous1946}, we could enhance the spontaneous emission rate without affecting the other parameters such as $g$ and $\gamma_c$ by reducing the mode-volume of the cavity. In practice, micro- or nano-cavities such as photonic crystal cavities will serve this purpose \cite{Vahala2003,Birowosuto2012,Takiguchi2013z,Sumikura2016,Ren2018,Mork2018}.
For a $\beta$ smaller than unity ($\beta<1$), we need to consider a finite $\gamma_A$. When $\beta\ll1$, the Purcell enhancement of the carrier decay rate is negligible and $\gamma_\|\simeq\gamma_A$ holds. Thus, the conditions  Eq. (\ref{condition0})-(\ref{condition3}) will be summarized as
\begin{equation}
\gamma_\perp\gg\gamma_A\gg\gamma_c\ \ {\rm and}\ \gamma_A\gamma_\perp\gg g^2\ \ {\rm for}\ \beta\ll1.
\end{equation}
Therefore, for each $\beta$, we need to choose an adequate set of parameters $\gamma_\perp$, $\gamma_A$, and $g$. However, since these parameters are absorbed in $\beta$ and $\gamma_\|$, they do not explicitly appear in the following discussion. 

Additionally, in realistic lasers, the lower level population is not negligible. Its effect is conventionally represented by replacing the stimulated emission term $Nn$ with $(N-N_0)n$ in the rate equations,   Eq. (\ref{eq:REp}) and (\ref{eq:REn}) \cite{Takemura2019}. In semiconductor lasers, the number $N_0$ is referred to as a carrier transparency number. To achieve high-$\beta$ class-A lasers, an additional condition $\beta N_0\ll 1$ is required because a finite $N_0$ works as  photon absorption and effectively reduces the cavity photon lifetime. We note that this condition regarding the carrier transparency number is practically very important because we cannot neglect $N_0$ in semiconductor lasers, and this is addressed in \cite{Takemura2019}. In fact, realizing $N_0\simeq0$ may be more difficult than the class-A condition because $\gamma_\parallel\gtrsim10\gamma_c$ is enough for the class-A regime, while four-level atoms with controlled decay rates are required for $N_0\simeq0$. 

In summary, high-$\beta$ class-A lasers with $N_0\simeq0$ are rather ideal cases and, in this paper, we study this regime with purely theoretical interests. However, we show that the role of $\beta$ becomes the most transparent in this regime.

\section{Birth-death master equation}
\subsection{Static photon statistics}
Now, we attempt to build a master equation based on the rate equation,  Eq. (\ref{eq:re_classA}). Since the first and the second terms on the RHS of Eq. (\ref{eq:re_classA}) represent the photon annihilation and creation processes, respectively, the master equation may read
\begin{eqnarray}
\dot{p}_{n}&=&-\gamma_cnp_{n}+\gamma_c(n+1)p_{n+1}\nonumber\\
&&-\frac{\beta(n+1)P}{\beta n+1} p_{n}+\frac{\beta nP}{\beta (n-1)+1} p_{n-1},\label{eq:master1}
\end{eqnarray}
where $p_{n}$ is the probability of finding $n$ photons inside the cavity. The probability $p_{n}$ is equivalent to the diagonal parts of the density matrix of the photons:  $p_{n}=\rho_{n,n}$. Here, the density matrix $\rho_{n,n'}$ is the expansion coefficient of the density operator of the system expressed as $\hat{\rho}(t)=\sum_{n,n'}\rho_{n,n'}(t)|n\rangle\langle n'|.$ The master equation,  Eq. (\ref{eq:master1}), has the same form as the diagonal parts of the Scully-Lamb master equation described in Section 5. Importantly, Eq. (\ref{eq:master1}) reproduces the pump-input and light-output curve predicted by the conventional rate equations,  Eqs. (\ref{eq:REp}) and (\ref{eq:REn}). It is worth noting that the spontaneous emission coupling coefficient $\beta$ has already been introduced in the class-A laser master equation in a few studies \cite{Jin1994,Loudon1973,Yamamoto1983}, but they are slightly different from ours. When $\beta$ is sufficiently low, all these master equations coincide with ours.

Although the direct numerical integration of Eq. (\ref{eq:master1}) is not computationally tough, we start from an analytical photon distribution. Let us consider the steady state $\dot{p}_{n}=0$. The detailed balance condition leads to
\begin{eqnarray}
-\gamma_cnp_{n}^{\rm ss}+\frac{\beta nP}{\beta(n-1)+1}p_{n-1}^{\rm ss}=0,\label{eq:detailed_balance}
\end{eqnarray}
where $p_n^{\rm ss}$ denotes the photon distribution in a steady state. We stress that the adiabatic elimination of the carrier (atomic) degree of freedom transforms the non-equilibrium laser system to an ``equilibrium" description satisfying the detailed balance condition Eq. (\ref{eq:detailed_balance}). From this equation, $p_n^{\rm ss}$ is derived as
\begin{eqnarray}
p_n^{\rm ss}=p_0^{\rm ss}\prod_{k=1}^{n}\frac{\tilde{P}}{k-1+1/\beta}
=Z^{-1}\frac{(1/\beta-1)!}{(n+1/\beta-1)!}\tilde{P}^n.\nonumber\\\label{eq:Pn}
\end{eqnarray}
With the normalization condition $\sum_{n=0}^\infty p_n^{\rm ss}=1$, the normalization coefficient $Z\equiv1/p_0^{\rm ss}$ is given by
\begin{eqnarray}
Z=\sum_{n=0}^\infty\frac{(1/\beta-1)!}{(n+1/\beta-1)!}\tilde{P}^n
=_1F_1(1;1/\beta;\tilde{P}),\label{eq:p0}
\end{eqnarray}
where $_1F_1$ is the confluent hypergeometric function and the factorial is defined in terms of the Gamma function as $x!=\Gamma(x+1)$. One important characteristic of Eq. (\ref{eq:p0}) is that the steady state photon distribution function $p_n^{\rm ss}$ depends only on two parameters: the spontaneous coupling coefficient $\beta$ and the normalized pump power $\tilde{P}$ [Eq. (\ref{eq:normpump})].

For conventional low-$\beta$ lasers where $\beta\ll1$, it is well known that as the pump power increases, a low-$\beta$ laser exhibits a transition from the thermal photon statistics $p_n=\left[1-(\beta \tilde{P})\right](\beta \tilde{P})^n$ to Poissonian (coherent) photon statistics $p_n= e^{-\tilde{P}}\tilde{P}^n/n!$. On the other hand, for $\beta=1$, regardless of the pump power, for any pump power $P$, the probability distribution function $p_n^{\rm ss}$ is given as
\begin{eqnarray}
p_n^{\rm ss}=e^{-\tilde{P}}\frac{\tilde{P}^n}{n!}\ \ {\rm for}\ \beta=1.\label{eq:coh}
\end{eqnarray}
Thus, a class-A ``laser" with $\beta=1$ emits Poissonian light at any pump rate. In fact, when $\beta=1$, the class-A rate equation takes a very simple form:
\begin{equation}
\dot{n}=-\gamma_cn+P\ \ {\rm for}\ \beta=1,
\end{equation}
where the Poissonian noise of the pump directly converts to photon statistics. Thus, the thresholdless ($\beta=1$) laser in the class-A limit is qualitatively different from conventional lasers. In Section 6, we discuss the experimental realization of class-A lasers with $\beta=1$. 

Now, we calculate various quantities using the photon distribution $p_n^{\rm ss}$ [Eq. (\ref{eq:Pn})]. In the beginning, the mean photon number $\langle \hat{n}\rangle$ is given by
\begin{eqnarray}
\langle \hat{n}\rangle&=&\sum_{n=0}^{\infty}np_n^{\rm ss}=\tilde{P}+(1-1/\beta)(1-1/Z),\label{eq:g2_1}
\end{eqnarray}
where we used the normalization condition $\sum_{n=1}^{\infty}p_n^{\rm ss}=1$ and the fact that $Z(=1/p_0^{\rm ss})$ is given by Eq. (\ref{eq:p0}). Similarly, the quantity $\langle\hat{n}(\hat{n}-1)\rangle$ is calculated as
\begin{eqnarray}
\langle \hat{n}(\hat{n}-1)\rangle&=&\sum_{n=2}^{\infty}(n-1)np_n^{\rm ss}\nonumber\\
&=&-(1/\beta)(1-1/\beta)(1-1/Z),\label{eq:g2_2}
\end{eqnarray}
where we use the relation $p_1^{\rm ss}=\beta p_0^{\rm ss}\tilde{P}$, which is inferred from Eq. (\ref{eq:Pn}). With these values, the second-order photon correlation function with a zero time delay $g^{(2)}(0)$ is given by
\begin{equation}
g^{(2)}(0)=\frac{\langle\hat{n}(\hat{n}-1)\rangle}{\langle\hat{n}\rangle^2}.\label{eq:g2_f}
\end{equation}
Figure \ref{fig:beta}(a) and (b) show $\langle n\rangle$ and $g^{(2)}(0)$ as a function of normalized pump power $P/P_{\rm th}$, respectively.
We note that the pump-input and light-output curve given by Eq. (\ref{eq:g2_1}) [see Fig. \ref{fig:beta}(a)] slightly deviates from that given by Eq. (\ref{eq:nsteady}) around the lasing threshold, which is interpreted as an effect of quantum fluctuations (quantum correlations), which is not taken into account in Eq. (\ref{eq:nsteady}) \cite{Meystre2007}. However, we also note that this difference may be too small to experimentally observe.

As we expect from the above discussion, Fig. \ref{fig:beta}(b) shows that a low-$\beta$ laser with $\beta$=0.001 presents a clear transition from thermal $g^{(2)}(0)=2$ to Poissonian photon statistics $g^{(2)}(0)=1$ at the threshold $P=P_{\rm th}$. In contrast, for a thresholdless laser with $\beta=1$, $g^{(2)}(0)$ is always unity, $g^{(2)}(0)=1$, because it always emits Poissonian light. When $\beta$ is high but not unity, for example when $\beta=0.5$, $g^{(2)}(0)$ reaches a constant value between 1 and 2 at a low pump power limit [see the magenta line in Fig. \ref{fig:beta}(b)]. To calculate $g^{(2)}(0)$ at a low pump power limit $P\rightarrow0$, we used a series expansion of the confluent hypergeometric function. At the low pump power limit, equations (\ref{eq:g2_1})-(\ref{eq:g2_f}) give rise to
\begin{eqnarray}
g^{(2)}(0)=\frac{\frac{2\beta^2}{1+\beta}\tilde{P}^2+\mathcal{O}(\tilde{P}^3)}{\beta^2\tilde{P}^2+\mathcal{O}(\tilde{P}^3)}\rightarrow\frac{2}{1+\beta}\ \ \ {\rm for}\ {P}\rightarrow0.\nonumber\\\label{eq:g2lim}
\end{eqnarray}
Figure \ref{fig:beta}(d) shows the low pump power limit of $g^{(2)}(0)$ as a function of $\beta$ based on Eq. (\ref{eq:g2lim}). It is clearly shown that high-$\beta$ class-A lasers emit partially coherent (Poissonian) light $1<g^{(2)}(0)<2$ at the low pump power limit.

\subsection{Dynamic photon statistics}
Next, we calculate the time delay dependence of the second-order photon correlations $g^{(2)}(\tau)$. For time independent systems, the time delay dependent second-order correlation $g^{(2)}(\tau)$ is defined as \cite{Verger2006}
\begin{eqnarray}
g^{(2)}(\tau)\equiv G^{(2)}(\tau)/\langle\hat{n}\rangle^2,\label{eq:defg2}
\end{eqnarray}
where
\begin{eqnarray}
G^{(2)}(\tau)&=&{\rm tr}\lbrace\hat{\rho}\hat{a}^\dagger(0)\hat{a}^\dagger(\tau)\hat{a}(\tau)\hat{a}(0)\rbrace\nonumber\\
&=&{\rm tr}\lbrace\hat{U}(\tau,0)\hat{a}\hat{\rho}(0)\hat{a}^\dagger\hat{U}^\dagger(\tau,0)\hat{a}^\dagger\hat{a}\rbrace.
\end{eqnarray}
Here, $\hat{U}(\tau,0)$ is a time evolution operator from time $t=0$ to $\tau$. If the single photon annihilation operation $\hat{\rho}(0)\rightarrow\hat{a}\hat{\rho}(0)\hat{a}^\dagger$ at time $t=0$ is interpreted as a perturbation that drives the system out of equilibrium, the function $G^{(2)}(\tau)$ represents the relaxation process of a perturbed system to a steady-state.
\begin{figure}
\centering
\includegraphics[width=0.35\textwidth]{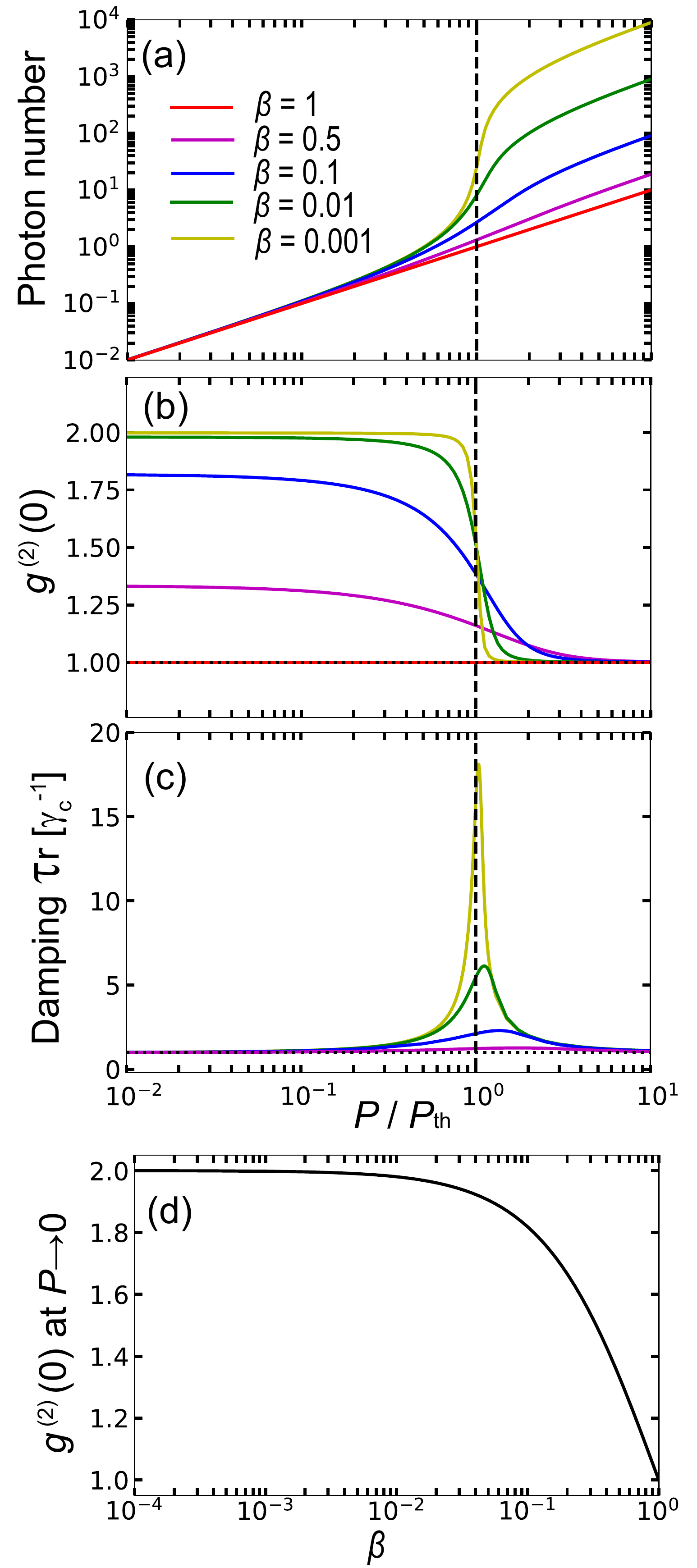}
\caption{(a) Output photon number $\langle\hat{n}\rangle$, (b) the second-order photon correlation at a zero time delay $g^{(2)}(0)$, and (c) the damping time of the second-order photon correlation $\tau_r$ calculated as a function of normalized pump power $P/P_{\rm th}$. Red, magenta, blue, green, and yellow lines, respectively, represent $\beta=1$, 0.5, 0.1, 0.01, and 0.001. (d) The low pump power limit ($P\rightarrow0$) of $g^{(2)}(0)$ is shown as a function of $\beta$, which is given by Eq. (\ref{eq:g2lim}).}
\label{fig:beta}
\end{figure}

Since in class-A lasers, $g^{(2)}(\tau)$ monotonically decays as the time delay $\tau$ increases, we performed exponential fitting: $g^{(2)}\propto\exp{(-|\tau|/\tau_r)}$. This monotonic decay is in contrast to $g^{(2)}(\tau)$ in class-B lasers, where the damped oscillatory behavior of $g^{(2)}(\tau)$ originates from the relaxation oscillation \cite{Takemura2012,Wang2015}. In Fig. \ref{fig:beta}(c), we show damping time $\tau_r$ as a function of pump power. For a low-$\beta$ laser with $\beta=0.001$, we find the enhancement of $\tau_r$ at the threshold $P=P_{\rm th}$. In terms of the analogy between the second-order phase transition and lasing, the enhancement of $\tau_r$ at the threshold is interpreted as the ``critical slowing down" of the amplitude mode; that is, the dynamics of the system are dramatically slowed at a critical point. Meanwhile, for a high-$\beta$ laser with $\beta=0.5$, the damping time $\tau_r$ is almost constant and independent of the pump power. In the next section, we argue that the suppression of the critical slowing down in high-$\beta$ lasers may be viewed as a breakdown of the laser-phase transition analogy in a ``small" system. When $\beta=1$, $\tau_r=1/\gamma_c$ will hold for any pump power, but we cannot extract $\tau_r$ because $g^{(2)}(\tau)$ becomes equal to one.
Note that since the damping times of second-order photon correlations $\tau_r$ shown in Fig. \ref{fig:beta}(c) characterize the relaxation times, they also approximate the turn-on times of the lasers (not shown).

\section{Coherence: Full Master Equation}
In the earlier sections, we investigated the photon statistical properties of class-A lasers using the birth-death master equation,  Eq. (\ref{eq:master1}), which represents the dynamics of the diagonal part of the photon density matrix. Unfortunately, Eq. (\ref{eq:master1}) does not provides information on laser linewidth (the first-order coherence), which is associated with the off-diagonal part of the photon density matrix. Since a full density matrix is necessary, the calculation of laser linewidth is a challenging task. One approach is to directly simulate an ensemble of four-level atoms that couples with cavity photons using the Lindblad type quantum master equations. However, these simulations are computationally too demanding and still limited to a few atoms due to a huge Liouville space. One alternative approach might be to simulate the Heisenberg equations of motion of higher-order correlations with the aid of truncation techniques \cite{Gies2007,Gartner2019}. Here, instead, we employ the well-known master equation theory developed by Scully and Lamb to describe lasers \cite{Scully1966,Scully1967,Scully1999}. Although the Scully-Lamb model is very simple, it depicts most of the important properties of lasing transitions such as the intensity jump, photon statistics, and linewidth. In this section, we introduce the spontaneous coupling coefficient $\beta$ into the Scully-Lamb model.

\subsection{Scully-Lamb quantum theory of lasers}
Following \cite{Scully1967}, we briefly review the Scully-Lamb quantum theory of lasers. Figure \ref{fig:rate_eq}(b) shows a schematic of the Scully-Lamb model. We consider two-level atoms injected into a cavity at a rate $r_a$. The injected atoms interact with photons inside the cavity with a coupling constant $g$. The decay rate of photons from the cavity is represented by $\gamma_c$. The upper $a$ and lower level $b$ of the two levels of atoms have decay rates of $\gamma_a$ and $\gamma_b$, respectively. The transverse relaxation rate of the two-level atoms, $\gamma_{ab}$, is assumed to be $\gamma_{ab}=(\gamma_a+\gamma_b)/2$. Importantly, the lifetime of photons inside the cavity must be much longer than the interaction time between the atoms and cavity photons, which results in tracing out the atomic degree of freedom (class-A condition). We return to the detailed requirements needed to satisfy this condition. For simplicity, we assume a resonance between the atomic transition and cavity (zero detuning). With these assumptions, in a coarse-grained time derivative, the non-Lindbladian master equation of the photons on a photon number basis is given by \cite{Scully1966,Scully1967}
\begin{eqnarray}
&&\dot{\rho}_{n,n'}=-\frac{\gamma_c}{2}(n+n')\rho_{n,n'}+\gamma_c\sqrt{(n+1)(n'+1)}\rho_{n+1,n'+1}\nonumber\\
&&-\frac{g^2\gamma_b\gamma_{ab}(n+1+n'+1)+g^4(n-n')^2}{\gamma_a\gamma_b\gamma_{ab}^2+2g^2\gamma_{ab}^2(n+1+n'+1)+g^4(n-n')^2}\ r_a\rho_{n,n'}\nonumber\\
&&+\frac{2g^2\gamma_b\gamma_{ab}\sqrt{nn'}}{\gamma_a\gamma_b\gamma_{ab}^2+2g^2\gamma_{ab}^2(n+n')+g^4(n-n')^2}\ r_a\rho_{n-1,n'-1}.\nonumber\\\label{eq:Scully-Lamb}
\end{eqnarray}
The diagonal part of the density matrix $n=n'$ reads
\begin{eqnarray}
\dot{\rho}_{n,n}&=&-\gamma_cn\rho_{n,n}+\gamma_c(n+1)\rho_{n+1,n+1}\nonumber\\
&&-\frac{2g^2\gamma_b\gamma_{ab}(n+1)}{\gamma_a\gamma_b\gamma_{ab}+4g^2\gamma_{ab}^2(n+1)}r_a\rho_{n,n}\nonumber\\
&&+\frac{2g^2\gamma_b\gamma_{ab} n}{\gamma_a\gamma_b\gamma_{ab}+4g^2\gamma_{ab}^2n}r_a\rho_{n-1,n-1}.\label{eq:Scully-Lamb_diag}
\end{eqnarray}

Eq. (\ref{eq:Scully-Lamb_diag}) is reduced to Eq. (\ref{eq:master1}). It is worth noting that the first line on the RHS of Eq. (\ref{eq:Scully-Lamb}) represents the conventional Lindblad-type photon decay term \cite{Meystre2007}, which is based on a second-order system-reservoir interaction and the Markov approximation. On the other hand, the second and third lines on the RHS of Eq. (\ref{eq:Scully-Lamb}) have an infinite order of interaction between the atomic reservoir and the photons \cite{Mandel1995,Henkel2007}, which is inferred from the fact that the coupling constant $g$ is present in the denominator. Unlike the Lindblad form, the Scully-Lamb master equation (\ref{eq:Scully-Lamb}) was heuristically derived, and its operator representation is not known \footnote{The operator representation of the fourth-order perturbation of the Scully-Lamb master equation is known \cite{Mandel1995,Yamamoto1999,Henkel2007,Arkhipov2019}. However, the perturbative Scully-Lamb master equation is applicable only for the vicinity of the lasing threshold of low-$\beta$ lasers.}.

Importantly, the diagonal part of the Scully-Lamb master equation (\ref{eq:Scully-Lamb_diag}) has the same form as the birth-death master equation (\ref{eq:master1}). Therefore, comparing Eq. (\ref{eq:Scully-Lamb_diag}) and (\ref{eq:master1}), the spontaneous emission coupling coefficient $\beta$ and pump $P$ are introduced as
\begin{equation}
\beta=\frac{4g^2/\gamma_b}{\gamma_a+4g^2/\gamma_b}\label{eq:beta_scully}
\end{equation}
and
\begin{equation}
r_a=\frac{2\gamma_{ab}}{\gamma_b}P,\label{eq:scully-pump}
\end{equation}
respectively. In order to introduce $\beta$, we compare Eq. (\ref{eq:beta}) and (\ref{eq:beta_scully}). 

\subsection{Conditions for high-$\beta$ class-A lasers in the Scully-Lamb model}
We discuss conditions for high-$\beta$ class-A lasers in the Scully-Lamb model in the same way as in Section 2D. We notice that the decay rate of the upper level $\gamma_a$ is analogous to $\gamma_A$. Namely, $\gamma_a$ represents the decay rate of the upper level population into non-lasing modes. To consider the conditions required for the Scully-Lamb model, we introduce the effective decay rate of the upper level population defined as
\begin{equation}
\gamma_{\rm eff}=\gamma_a+\frac{4g^2}{\gamma_c+\gamma_{ab}}.
\label{eq:LS_geff}
\end{equation}
This effective decay rate is analogous to $\gamma_\|$ in the rate equation approach and can be derived by using the Jaynes-Cummings model with dephasing \cite{Auffeves2009}. The second part of the RHS of Eq. (\ref{eq:LS_geff}) represents the Purcell enhancement of the population decay. Since the Scully-Lamb model assumes a weak coupling condition and that the atom dynamics are much faster than the photon dynamics, the required conditions are
\begin{equation}
\gamma_{\rm eff}\gg\gamma_c.\ \ {\rm (class\mathchar`-A}\ {\rm limit)}.
\label{eq:LS_condition1}
\end{equation}
and
\begin{equation}
\gamma_{ab}>g\ \ {\rm (weak}\ {\rm coupling)}.
\label{eq:LS_condition2}
\end{equation}
First, we consider the conditions required for thresholdless lasing $\beta=1$, which is again realized by setting $\gamma_a=0$. Since $\gamma_a=0$, the upper level population decays only through the cavity via the atom-photon interaction $g$. When $\gamma_{ab}=\gamma_b/2$ is taken into account and  $\gamma_c\ll\gamma_{ab}=\gamma_b/2$ is assumed, the effective decay rate of the upper level population $\gamma_{\rm eff}\simeq8g^2/\gamma_b$ holds. Thus, the two conditions, Eq (\ref{eq:LS_condition1}) and (\ref{eq:LS_condition2}), can be summarized as
\begin{equation}
\gamma_b^2>g^2\gg\gamma_c\gamma_b\ \ {\rm for}\ \beta=1,\label{eq:condition_scully}
\end{equation}
which is analogous to Eq. (\ref{condition_rate}).

In the following simulations, we use fixed values of $\gamma_c=0.0001g$ and $\gamma_b=10g$, which satisfy Eq. (\ref{eq:condition_scully}) for $\beta=1$. Furthermore, in the simulations of the Scully-Lamb model, we vary $\beta$ by changing the value of $\gamma_a$. If the two conditions, Eq (\ref{eq:LS_condition1}) and (\ref{eq:LS_condition2}), are automatically satisfied for a $\beta$ lower than unity ($\beta<1$), because of the increase of $\gamma_{\rm eff}$ and $\gamma_{ab}$. 

\subsection{Steady state}
First, let us discuss the steady state property of the Scully-Lamb master equation. Note that the diagonal part of the Scully-Lamb master equation (\ref{eq:Scully-Lamb_diag}) is the closed equation of motion, and it is identical to the birth-death master equation (\ref{eq:master1}) by introducing $\beta$ and $P$. Therefore, if the initial state is the vacuum state $\hat{\rho}_{\rm ini}=|0\rangle\langle0|$, the steady state density matrix should have non-zero values only in the diagonal parts. Even if the density matrix of the initial state has finite off-diagonal parts, they decay in time \cite{Scully1968}. Based on the discussion in Section 3, the steady state density operator  is given by
\begin{equation}
\hat{\rho}^{\rm ss}=\sum_{n=0}^{\infty}p_n^{\rm ss}|n\rangle\langle n|,\label{eq:equi}
\end{equation}
where $p_n^{\rm ss}$ is the steady state probability distribution given analytically by Eq. (\ref{eq:Pn}). Since the master equation,  Eq. (\ref{eq:Scully-Lamb}), is $U(1)$ symmetric and does not have a preferred phase \cite{Gartner2018}, the zero off-diagonal elements in the steady state density matrix are  natural.

More specifically, far above the lasing threshold ($P\gg P_{\rm th}$), since $p_n^{\rm ss}$ becomes Poissonian, the steady state density matrix is written as
\begin{equation}
\hat{\rho}^{\rm ss}=\sum_{n=0}^{\infty}e^{-\tilde{P}}\frac{\tilde{P}^n}{n!}|n\rangle\langle n|=\int_0^{2\pi}\frac{d\theta}{2\pi}||\alpha|e^{i\theta}\rangle\langle|\alpha|e^{i\theta}|,\label{eq:mix}
\end{equation}
where $||\alpha|e^{i\theta}\rangle=|\alpha\rangle=e^{-|\alpha|/2}\sum_{n=0}^{\infty}{\alpha^n}/{\sqrt{n!}}|n\rangle$ is the coherent state with $|\alpha|^2=\tilde{P}$. Equation (\ref{eq:mix}) is a mixed state of coherent states with phases varying from 0 to $2\pi$ and has phase symmetry [$U(1)$ gauge symmetry] \cite{Gartner2018}. Thus, without a symmetry breaking field, the Scully-Lamb master equation does not give rise to spontaneous symmetry breaking. However, in Section 5, we discuss the possibility of symmetry breaking in a ``thermodynamic limit" from the standpoint of the Liouvillian gap.

\subsection{Linewidth}
Here, we calculate the laser linewidth based on the Scully-Lamb master equation. Note that we skip the discussion of the photon statistical properties of the Scully-Lamb model because they are the same as in Figs. \ref{fig:beta}.

With the Wiener-Khintchine theorem, a spectral function $S(\nu)$ is given by \cite{Meystre2007} $S(\nu)=\int_{0}^{\infty}G^{(1)}(\tau)e^{-\nu\tau}d\tau,$ where $G^{(1)}(\tau)$ is the first-order photon correlation function defined as \cite{Scully1968}
\begin{eqnarray}
G^{(1)}(\tau)&=&\langle\hat{a}^\dagger(\tau)\hat{a}(0)\rangle={\rm tr}\lbrace\hat{\rho}\hat{a}^\dagger(\tau)\hat{a}(0)\rbrace\nonumber\\
&=&{\rm tr}\lbrace\hat{U}(\tau,0)\hat{a}\hat{\rho}(0)\hat{U}^\dagger(\tau,0)\hat{a}^\dagger\rbrace
\end{eqnarray}
Therefore, the decay rate of the first-order photon correlation function represents the laser linewidth. In general, the simulation of the master equation, Eq. (\ref{eq:Scully-Lamb}), with an $N$ photon basis requires the integration of $N\times N$ coupled equations, which is computationally tough, but a closed equation of motion for the off-diagonal density matrix greatly reduces the dimension $N\times N$ to the order of $N$. First, for the steady-state [see Eq. (\ref{eq:equi})], the operation at $t=0$ is expressed as
\begin{eqnarray}
\hat{\rho}(0)\rightarrow\hat{a}\hat{\rho}(0)&=&\sum_{n}\sqrt{n}p_{n}^{\rm ss}|n-1\rangle\langle n|
\end{eqnarray}
Second, importantly, the Scully-Lamb master equation couples only elements parallel to the diagonal elements. Thus, a closed equation of motion for ${\rho}_{n,n+1}$ is given by
\begin{eqnarray}
\dot{\rho}_{n,n+1}&=&-\frac{\gamma_c}{2}(2n+1)\rho_{n,n+1}\nonumber\\
&&+\gamma_c\sqrt{(n+1)(n+2)}\rho_{n+1,n+2}\nonumber\\
&&-\frac{g^2\gamma_b\gamma_{ab}(2n+3)+g^4}{\gamma_a\gamma_b\gamma_{ab}^2+2g^2\gamma_{ab}(2n+3)+g^4}r_a\rho_{n,n+1}\nonumber\\
&&+\frac{2g^2\gamma_b\gamma_{ab}\sqrt{n(n+1)}}{\gamma_a\gamma_b\gamma_{ab}^2+2g^2\gamma_{ab}^2(2n+1)+g^4}r_a\rho_{n-1,n}.\nonumber\\\label{eq:off-diag}
\end{eqnarray}
Since Eq. (\ref{eq:equi}) and (\ref{eq:off-diag}) involve only an $N$ photon basis rather than an $N\times N$ density matrix, numerical simulations of the first-order correlation $G^{(1)}(\tau)$ become easy. The laser linewidth $\Delta\nu$ is obtained as the decay rate of the first-order photon correlation: $G^{(1)}(\tau)\simeq G^{(1)}(0)e^{-\Delta\nu\tau}$.

\begin{figure}
\centering
\includegraphics[width=0.35\textwidth]{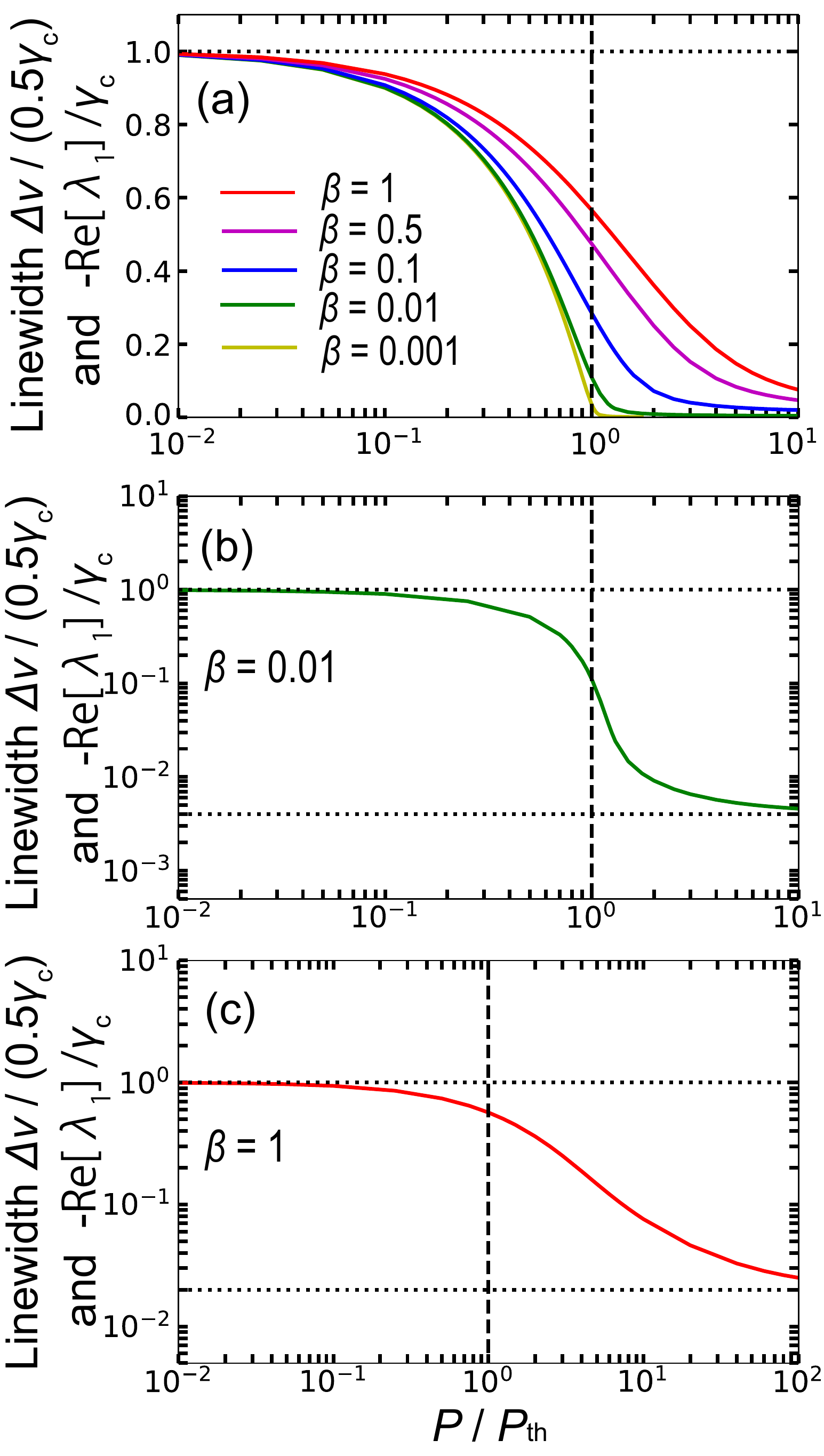}
\caption{(a) Laser linewidth with different $\beta$ as a function of normalized pump power $P/P_{\rm th}$. The line colors mean the same as in Fig. \ref{fig:beta}. The parameters used for the simulations are $\gamma_c=0.0001g$, and $\gamma_b=10g$. $\gamma_a$ is changed with Eq. (\ref{eq:beta_scully}). (b) and (c) Laser linewidths for $\beta=0.01$ and 1, respectively. The horizontal dotted lines represent the low- and high-power limits of the laser linewidth given by Eqs. (\ref{eq:low}) and (\ref{eq:high}). Linewidths shown in (a-c) are equivalent to the Liouvillian gaps $-{\rm Re}[\lambda_1]$ for corresponding $\beta$s.}
\label{fig:linewidth}
\end{figure}

In Fig. \ref{fig:linewidth}(a), we show numerically simulated laser linewidth as a function of the normalized pump power for five different $\beta$. Interestingly, even the thresholdless laser ($\beta=1$) exhibits the linewidth narrowing. Therefore, we may prove ``lasing transition" of class-A thresholdless lasers by measuring the linewidth of emission. In Fig. \ref{fig:linewidth}, panels (a) and (b) show linewidths in a log-log scale as a function of the normalized pump power $P/P_{\rm th}$ for $\beta=0.01$ and 1, respectively. They clearly show the linewidth decrease with increases in pump power. Taking a close look at the variation of the linewidth, we find that both in low ($P\rightarrow0$) and high pump power limit ($P\rightarrow\infty$), the linewidth asymptotically approaches finite values. In order to understand the linewidth behavior in these two limits, we analytically calculate the off-diagonal part of the photon density matrix following the discussion in \cite{Scully1967}. We assume the dynamics of $\rho_{n,n+1}$ as
\begin{equation}
\rho_{n,n+1}(t)=e^{-D_n(t)}\rho_{n,n+1}(0).
\label{eq:nu_assum}
\end{equation}
In the lowest order, $D_n(t)$ is approximated as $D_n(t)=\mu_nt$, where $\mu_n$ is given by
\begin{eqnarray}
&&\mu_n\simeq\nonumber\\
&&\frac{g^2\gamma_b\gamma_{ab}(2n+3)+g^4-2g^2\gamma_b\gamma_{ab}\sqrt{(n+1)(n+2)}}{\gamma_a\gamma_b\gamma_{ab}^2+2g^2\gamma_{ab}^2(2n+3)+g^4}r_a\nonumber\\
&&+\frac{1}{2}\gamma_c(2n+1)-\gamma_c\sqrt{n(n+1)}.\label{eq:mu}
\end{eqnarray}
In the low pump power limit ($r_a\rightarrow0$ and $n\rightarrow0$), the first part of Eq. (\ref{eq:mu}) becomes negligible and $\mu_n$ approaches $\gamma_c/2$, which leads to
\begin{equation}
\Delta\nu\rightarrow\frac{\gamma_c}{2}\ \ {\rm for}\ P\rightarrow0.\label{eq:low}
\end{equation}
Thus, far below the lasing threshold, the linewidth is determined solely by the cavity Q value. Meanwhile, in the high pump power limit ($r_a\rightarrow\infty$ and $n\rightarrow\infty$), the second part of Eq. (\ref{eq:mu}) is negligible and the linewidth behaves as
\begin{eqnarray}
\Delta\nu&\rightarrow&\frac{g^2r_a}{4\gamma_{ab}^2\langle n\rangle}\label{eq:high0}\\
&\rightarrow&\frac{g^2\gamma_c}{2\gamma_{ab}\gamma_b}\ \ {\rm for}\ P\rightarrow\infty.\label{eq:high}
\end{eqnarray}
To derive the last formula, we used Eq. (\ref{eq:scully-pump}) and the relation $\langle n\rangle\simeq \tilde{P}$ far above the lasing threshold. Eq. (\ref{eq:high0}) resembles the famous Schawlow-Townes linewidth $\Delta\nu\propto\langle n\rangle^{-1}$ \cite{Schawlow1958}, which indicates that the linewidth is inversely proportional to the output photon number in the high-pump power limit. However, we found that, in the Scully-Lamb model, the linewidth narrowing is limited, which could be due to the gain noise introduced by the pump. The dashed lines in Fig. \ref{fig:linewidth} are the low and high pump power limits of the laser linewidth given by Eq. (\ref{eq:low}) and (\ref{eq:high}).
We also note  Eq. (\ref{eq:high0}) indicates that the linewidth reaches zero ($\Delta\nu=0$) in the limit $\beta\rightarrow0$, which is achieved with $g\rightarrow0$ or $\gamma_a\rightarrow\infty$.

\section{Laser-phase transition analogy}
In this section, we discuss the impact of $\beta$ on the laser-phase transition analogy \cite{DeGiorgio1970,Graham1970,Gartner2018}. The relation between $\beta$ and effective system size was first discussed by Rice and Carmichael, but not in the class-A limit. We argue that this relation becomes clear in the class-A limit. Intuitively, this is because the class-A laser is effectively described as an equilibrium system satisfying the detailed balance condition (see Section 3A). For the amplitude mode (the diagonal part of the photon density matrix), photon statistics of a class-A laser are uniquely characterized by $\beta$. Meanwhile, for class-B lasers, the photon statistics are neither determined by $\beta$ nor by the ratio $\gamma_\|/\gamma_c$. 

As explained in \cite{Rice1994,Carmichael2015}, the effective ``system size" for nonlinear photonic systems is associated with the photon number required to activate nonlinearity. In our system, the number of photons at the lasing threshold $P=P_{\rm th}$ is given by $n_{\rm th}={\beta}^{-1/2}$ \cite{Rice1994,Druten2000,Elvira2011}. The number $n_{\rm th}$ could be interpreted as the photon number required to activate the nonlinearity of the system, and thus the ``system size" could be characterized by ${\beta}^{-1/2}$. With this definition, the ``thermodynamic limit" where the ``system size" is infinite ($n_{\rm th}\rightarrow\infty$) is the limit ${\beta}\rightarrow 0$. In fact, in the limit $\beta\ll0$, the nonlinearity is weak enough for the perturbative expansion of the master equation, Eq. (\ref{eq:master1}), up to the second-order of $\beta$ ($4^{\rm th}$ order of $g$). Furthermore, in the limit ${\beta}\rightarrow 0$, the spontaneous emission can be treated as perturbation; actually, as Langevin noises. In this regime, we can find a clear analogy between lasing and the second-order phase transition. For example, we can define the ``Ginzburg-Landau free energy" \cite{DeGiorgio1970}. Meanwhile, when $\beta$ approaches unity $\beta\rightarrow1$, the perturbative expansion of the master equation fails because the nonlinearity is too strong and all higher-order atom-photon interactions must be considered. In this sense, high-$\beta$ class-A lasers are operating in a non-perturbative regime, and the master equations,  Eq. (\ref{eq:master1}) and Eq. (\ref{eq:Scully-Lamb}), are the non-perturbative models that implicitly include all higher-order atom-photon interactions. In fact, we have already seen the suppression of the critical slowing down of $\tau_r$ in the small size limit $\beta\rightarrow1$ [See Fig. \ref{fig:beta}(c)], which is one of the signatures of the breakdown of the phase transition in the amplitude mode (diagonal part of the density matrix).

To discuss the laser-phase transition analogy in terms of both the amplitude and phase mode (off-diagonal part of the density matrix), we need to analyze the full density matrix Eq. (\ref{eq:Scully-Lamb}). For the analysis, we employ the Liouvillian gap and the Liouvillian spectrum theory, which clarify the nature of the dissipative phase transition with continuous phase symmetry. 

\subsection{Liouvillian gap}
We investigate the Liouvillian gap of the Scully-Lamb master equation of class-A lasers. It is known that the Liouvillian gap plays a central role in dissipative quantum phase transitions \cite{Kessler2012,Horstmann2013,Casteels2017}. Thus, it is currently being investigated in a wide variety of open quantum systems, including spin \cite{Kessler2012,Cai2013}, nonlinear optical systems \cite{Casteels2017,Rodriguez2017,Fink2017,Shirai2018}, and boundary time crystals \cite{Iemini2018}

In general, a quantum master equation for a general density operator $\hat{\rho}$ is written as
\begin{equation}
\frac{d}{dt}\hat{\rho}=\hat{\mathcal{L}}\hat{\rho},
\end{equation}
where $\hat{\mathcal{L}}$ is called a Liouvillian. We calculate the eigenvalues of the Liouvillian $\lambda_i$: $\hat{\mathcal{L}}\hat{\rho}_i=\lambda_i\hat{\rho}_i$. Here, we sort the eigenvalues as $-{\rm Re}[\lambda_{0}]<-{\rm Re}[\lambda_{1}]<\ldots<-{\rm Re}[\lambda_{i}]|$. In general, the first eigenvalue is zero ($\lambda_{0}=0$), and the first eigenmatrix $\hat{\rho}_0$ corresponds to the steady-state solution $\hat{\mathcal{L}}\hat{\rho}_0=\lambda_{0}\hat{\rho}_0=0$. Now, we can  define the ``gap" between the real parts of the first and second eigenvalues $|{\rm Re}[\lambda_1]-{\rm Re}[\lambda_0]|=-{\rm Re}[\lambda_1]$ of the Liovillian gap, which represents the smallest decay rate of the system (the critical decay rate) \cite{Minganti2018}. A dissipative phase transition occurs when the Liouvillian gap closes $-{\rm Re}[\lambda_1]\rightarrow0$ in a thermodynamic limit. We also note that the trace of eigenmatrix $\hat{\rho}_i$ for $i>0$ is not generally unity: $\mathrm{Tr}\hat{\rho}_{i(i>0)}\neq1 $; thus, $\hat{\rho}_i$ for $i>0$ is not a conventional density matrix.

The matrix elements of the Liouvillian of the Scully-Lamb master equation are given by Eq. (\ref{eq:Scully-Lamb}). In the rotating frame of the cavity frequency, the eigenvalues of the Liouvillian of the Scully-Lamb master equation are real ${\rm Im}[\lambda_i]=0$. First, since the first eigenmatrix $\hat{\rho}_0$ represents the steady state, $\hat{\rho}_0$ is given by Eq. (\ref{eq:equi}) for the Scully-Lamb master equation. From numerical calculation, surprisingly, we found that the Liouvillian gap for the Scully-Lamb master equation is identical to the laser linewidth shown in Fig. \ref{fig:linewidth}(a): $-{\rm Re}[\lambda_1]=\Delta\nu$. The coincidence of the Liouvillian gap and laser linewidth can be intuitively understood as follows. Since the Liouvillian gap represents the longest relaxation of the system (the critical decay rate), in our system, it corresponds to the phase diffusion process that is associated with the linewidth. This point is discussed again in Section 5B.

Now, let us discuss the thermodynamic limit defined as $\beta\rightarrow0$ ($n_{\rm sat}\rightarrow\infty$) for the Scully-Lamb model. From Fig. \ref{fig:linewidth}(a), in the limit $\beta\rightarrow0$, we can expect that the Liouvillian gap (or laser linewidth) will drop sharply to zero at the lasing threshold $P=P_{\rm th}$ and remain closed in the entire region above the threshold. This is in striking contrast to the Liouvillian gap for the first-order dissipative phase transition induced by the Kerr nonlinearity and coherent pumping, where the Liouvillian gap has its minimum value around the critical point \cite{Casteels2017,Kessler2012}. Our observation can be understood as the Liouvillian gap for the second-order dissipative phase transition with broken symmetries, which was recently studied by Ref. \cite{Minganti2018}. In fact, the closure of the Liouvillian gap in the whole ordered phase coincides with the prediction in \cite{Minganti2018}.

Furthermore, Ref. \cite{Minganti2018} investigated symmetry breaking of a discrete $Z_n$ symmetry associated with the dissipative phase transition  from the standpoint of a Liouvillian eigenvalues spectrum. As we discussed above, the Liouvillian associated with the master equation (\ref{eq:Scully-Lamb}) has phase symmetry [continuous $U(1)$ gauge symmetry]. Therefore, in the next subsection, we  discuss $U(1)$ phase symmetry breaking in the Scully-Lamb model using a Liouvillian eigenvalues spectrum. 
\begin{figure*}[t]
\centering
\includegraphics[width=0.95\textwidth]{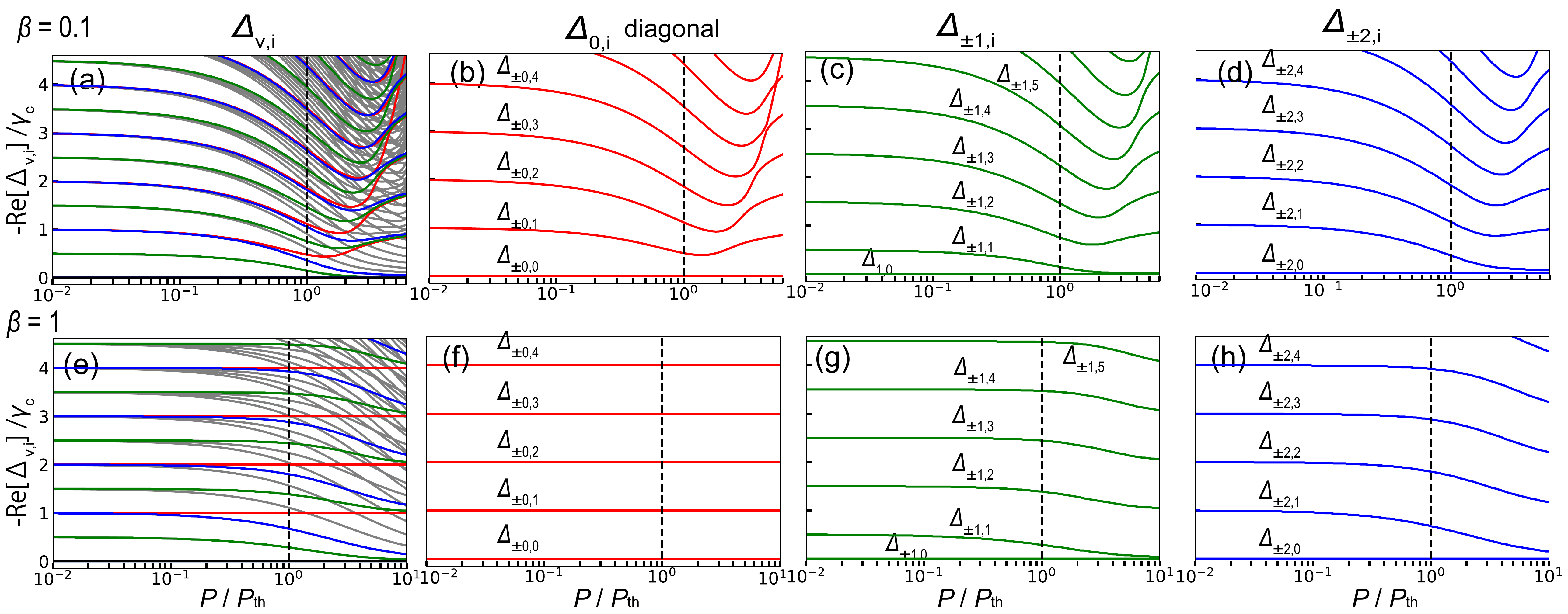}
\caption{(a), (e)  Liouvillian eigenvalue eigenvalue spectra. Liouvillian eigenvalue $\Delta_{0,i}$ (b), (f), $\Delta_{\pm1,i}$ (c), (g), and $\Delta_{\pm2,i}$ (d), (h). Panels (a-d) and (e-h) are respectively for $\beta=0.1$ and 1. The parameters are the same as in Fig. \ref{fig:linewidth}. }
\label{fig:spectrum}
\end{figure*}

\subsection{Liouvillian eigenvalue spectrum}
Figure \ref{fig:spectrum}(a) and (e) directly show several of the smallest eigenvalues of the Liouvillian for $\beta=0.1$ and $\beta=1$, respectively. Even though the spectra of the eigenvalues appear very complicated, in the following, we show that they can be systematically classified by taking the symmetry of the Liouvillian into account. As we mentioned earlier, the Scully-Lamb master equation couples only the density matrix elements parallel to the diagonal part. Therefore, the Scully-Lamb master equation,  Eq. (\ref{eq:Scully-Lamb}), can be decomposed into two triangles of the density matrix as
\begin{eqnarray}
\dot{\rho}_{n,n+\nu}=\sum_{m=0}^\infty{M}^{(\nu)}_{n,m}{\rho}_{m,m+\nu}\ \ {\rm for}\ \nu=0,1,2,\cdots
\label{eq:master_nu}
\end{eqnarray}
and
\begin{eqnarray}
\dot{\rho}_{n-\nu,n}=\sum_{m=\nu}^\infty{M}^{(\nu)}_{n,m}{\rho}_{m-\nu,m}\ \ {\rm for}\ \nu=-1,-2,\cdots.
\label{eq:master_nu2}
\end{eqnarray}
Here, $\nu$ is the index that designates the matrix elements parallel to the diagonal part. The ${M}^{(\nu)}$ for any $\nu$ is easily determined from Eq. (\ref{eq:Scully-Lamb}). Now, we consider the $i$th eigenvalue of the matrix ${M}^{(\nu)}$ denoted as $\Delta_{\nu,i}$ ($i=0,1,2\ldots$), where the eigenvalues are sorted as $-{\rm Re}[\Delta_{\nu,0}]<-{\rm Re}[\Delta_{\nu,1}]<\ldots<-{\rm Re}[\Delta_{\nu,i}]$. As described in Section, ${\rm Re}[\Delta_{-\nu,i}]={\rm Re}[\Delta_{\nu,i}]$ and ${\rm Im}[\Delta_{-\nu,i}]=-{\rm Im}[\Delta_{\nu,i}]$ hold, but, in our case, ${\rm Im}[\Delta_{\nu,i}]=0$ because we are using the rotating frame of the cavity. It is important to note that the eigenvalue $\Delta_{\nu,i}$ represents the decay rate of the the eigenmatrix ${\rho}_{n,n+\nu(\geq0)}$ or ${\rho}_{n-\nu(<0),n}$ into the steady state [eq. (\ref{eq:equi})]. For $\nu\neq0$, the steady state is a zero matrix.

We show the Liouvillian eigenvalue spectrum ($\Delta_{\nu,i}$) in Fig. \ref{fig:spectrum} for $\beta=0.1$ (a) and $1$ (e). Note that $\Delta_{\nu,0}=0$ holds for any $\nu$ because the first eigenvalue of the Liouvillian is $\lambda_0=0$ (recall the discussion in Section 5A). First, let us discuss $\nu=0$, which is associated with the diagonal part of the density matrix and represents the amplitude mode. The master equation for $\nu=0$: $\dot{\rho}_{n,n}=\sum_{m=0}^\infty{M}^{(0)}_{n,m}{\rho}_{m,m}$ is equivalent to the birth-death master equation given by Eq. (\ref{eq:Scully-Lamb_diag}) or (\ref{eq:master1}). Figure \ref{fig:spectrum}(b) and (f) show several smallest eigenvalues $-{\rm Re}[\Delta_{0,i}]$. Here, since $-{\rm Re}[\Delta_{0,1}]$ represents the critical decay of the amplitude mode, it approximately coincides with the decay rate of $g^{(2)}(\tau)$ shown in Fig. \ref{fig:beta}(c): $-{\rm Re}[\Delta_{0,1}]\simeq1/\tau_r$. For $\beta=0.1$, $-{\rm Re}[\Delta_{0,1}]$ shown in Fig. \ref{fig:spectrum}(b) has a minimum value around the lasing threshold, which corresponds to a maximum damping time of $g^{(2)}(\tau)$ [see Fig. \ref{fig:beta}(c)]. In contrast, for $\beta=1$, as we explained in Section 3, $-{\rm Re}[\Delta_{0,1}]=\gamma_c$ for any pump power.

Second, we discuss the master equation for $\nu\neq0$, which is associated with the phase mode. First, let us consider $\nu=\pm1$. Considering that the master equation for $\nu=1$ is equivalent to Eq. (\ref{eq:off-diag}) and that the relation $-{\rm Re}[\Delta_{1,1}]=-{\rm Re}[\Delta_{-1,1}]$ holds, we can find that the eigenvalue $-{\rm Re}[\Delta_{\pm1,1}]$ represents the laser linewidth $-{\rm Re}[\Delta_{\pm1,1}]=\Delta\nu$. Furthermore, importantly, since $-{\rm Re}[\Delta_{\pm1,1}]$ is the smallest non-zero real part of the eigenvalues of the Scully-Lamb Liouvillian, $-{\rm Re}[\Delta_{\pm1,1}]$ is nothing else but the Liouvillian gap: $-{\rm Re}[\Delta_{\pm1,1}]=-{\rm Re}[\lambda_1]$. Interestingly, the Liouvillian gap $-{\rm Re}[\lambda_1]$ has two-fold degeneration ($-{\rm Re}[\Delta_{1,1}]$ and $-{\rm Re}[\Delta_{-1,1}]$), which intuitively represents the clockwise and counter-clockwise directions of the phase diffusion.
Finally, we show the eigenvalue $-{\rm Re}[\Delta_{\pm2,i}]$ in Fig. \ref{fig:spectrum}(d) and (h), which behaves in a similar manner to $-{\rm Re}[\Delta_{\pm1,i}]$. Applying the same argument of Eqs (\ref{eq:nu_assum})-(\ref{eq:high}) to the master equation for $\nu=2$, we find that $-{\rm Re}[\Delta_{\pm2,1}]$ starts from $\gamma_c$ ($P\rightarrow0$) and reaches ${2g^2\gamma_c}/{(\gamma_{ab}\gamma_b)}$ in the high pump power limit ($P\rightarrow\infty$). In general, for $\nu\neq0$, the eigenvalue follows as $-{\rm Re}[\Delta_{\pm\nu,1}]\rightarrow\frac{\gamma_c}{2}|\nu|$ in the low pump power limit $P\rightarrow0$. Meanwhile, in the high pump power limit, the eigenvalue behaves as 
\begin{eqnarray}
-{\rm Re}[\Delta_{\pm\nu(\neq0),1}]\rightarrow\frac{g^2r_a}{4\gamma_{ab}^2\langle n\rangle}\nu^2
\rightarrow\frac{g^2\gamma_c}{2\gamma_{ab}\gamma_b}\nu^2\ \ {\rm for}\ P\rightarrow\infty.
\label{eq:Lgap_limit}
\end{eqnarray}
Similarly to Eq. (\ref{eq:high0}), $-{\rm Re}[\Delta_{\nu(\neq0),1}]$ becomes zero in the limit $\beta\rightarrow0$.

In summary, the Liouvillian eigenvalue spectrum contains information on both the amplitude and phase mode. The eigenvalues $\Delta_{0,1}$ and $\Delta_{\pm1,1}$, respectively, represent the decay rates of $g^{(2)}(\tau)$ and $g^{(1)}(\tau)$, which care experimentally measurable.

\subsection{Symmetry breaking}
Now, we discuss symmetry breaking from the standpoint of the Liouvillian spectrum. In terms of symmetry, the eigenmatrix of the Liouvillian associated with nonzero $\nu$ has $Z_{|\nu|}$ symmetry, which can be proved in the following way. First, for simplicity, we consider positive nonzero $\nu$. The $i^{\rm th}$ eigenmatrix of the Liouvillian for $\nu$ can generally be written as
\begin{equation}
\hat{\rho}_{i}^{(\nu)}=\sum_{n=0}^\infty c^{\nu,i}_{n}|n\rangle\langle n+\nu|\ \ {\rm for}\ \nu=0,1,2,\cdots
\end{equation}
and
\begin{equation}
\hat{\rho}_{i}^{(\nu)}=\sum_{n=\nu}^\infty c^{\nu,i}_{n}|n-\nu\rangle\langle n|\ \ {\rm for}\ \nu=-1,-2,\cdots,
\end{equation}
where the coefficient $c^{\nu,i}_{n}$ is determined as the $i$th eigenvector of the matrix ${M}^{(\nu)}$. The eigenmatrix of the Liouvillian $\hat{\rho}_{i}^{(\nu)}$ can also be written as
\begin{eqnarray}
\hat{\rho}_{i}^{(\nu)}=\hat{\rho}_{\rm diag}\hat{a}^\nu\ \ {\rm for}\ \nu=0,1,2,\cdots,\label{eq:sym1}
\end{eqnarray}
and
\begin{eqnarray}
\hat{\rho}_{i}^{(\nu)}=\hat{a}^{|\nu|}\hat{\rho}_{\rm diag}\ \ {\rm for}\ \nu=-1,-2,\cdots.\label{eq:sym2}
\end{eqnarray}
Here, $\hat{\rho}_{\rm diag}=\sum_{n}^{\infty}a_n|n\rangle\langle n|$ is an arbitrary diagonal matrix, where we can always find the coefficient $a_n$. Now, let us consider the phase rotation [the global $U(1)$ gauge transformation] $\hat{a}\rightarrow\hat{a}e^{i\theta}$. Importantly, since the matrix $\hat{\rho}_{\rm diag}$ is invariant under phase rotation [the $U(1)$ gauge symmetry], it is clear that $\hat{\rho}_{i}^{(\nu)}$ is transformed as $\hat{\rho}_{i}^{(\nu)}\rightarrow\hat{\rho}_{i}^{(\nu)}e^{i|\nu|\theta}$. This indicates that the $U(1)$ gauge symmetry is broken in $\hat{\rho}_{i}^{(\nu)}$ and that it  is invariant under a phase rotation of $2\pi/|\nu|$ degrees: $Z_{|\nu|}$ symmetry. In fact, Eq. (\ref{eq:master_nu}) and (\ref{eq:master_nu2}) correspond to the decomposition of the Liouvillian according to a symmetry sector described in Ref. \cite{Minganti2018}. Furthermore, we point out that the eigenvalue $\Delta_{\nu,i}$ is analogous to the eigenvalue of the Fokker-Planck equation for low-$\beta$ lasers as intensively investigated by Risken and Vollmer \cite{Risken1967,Risken1996,Haken2012} and Lax and Louisell \cite{Lax1967}. As with $\nu$ and $i$, in the Fokker-Planck equation for the polar coordinate, the eigenvalue has two indices that correspond to the phase and the radial degrees of freedom. Therefore, a similar discussion may be possible based on the Fokker-Plank equation approach, but only for low-$\beta$ lasers.

Finally, we discuss symmetry breaking in the thermodynamic limit from the standpoint of the Liouvillian spectrum. As we already discussed, since the thermodynamic limit $\beta\rightarrow0$ is achieved with $\gamma_a\rightarrow\infty$ or $g\rightarrow0$ [See Eq. (\ref{eq:beta_scully})], Eq. (\ref{eq:Lgap_limit}) indicates that the Liouvillian eigenvalue $-{\rm Re}[\Delta_{\nu(\neq0),1}]$ behaves as $-{\rm Re}[\Delta_{\nu(\neq0),1}]\rightarrow0$ in the high pump power limit. Furthermore, assuming that the Liouvillian eigenvalue $-{\rm Re}[\Delta_{\nu(\neq0),1}]$ will behave in the same way as the linewidth shown in Fig. \ref{fig:linewidth}, we expect the following:
\begin{eqnarray}
-{\rm Re}[\Delta_{\nu(\neq0),1}]\rightarrow0\ \ {\rm for}\ P>P_{\rm th}\ \ {\rm and}\ \ \beta\rightarrow0
\label{sym_breaking}
\end{eqnarray}
This equation indicates that in the thermodynamic limit ($\beta\rightarrow0$), the eigenvalue $-{\rm Re}[\Delta_{\nu(\neq0),1}]$ for any $\nu$ is asymptotically zero above the lasing threshold. Recalling that the Liouvillian eigenvalue $-{\rm Re}[\Delta_{\nu,1}]$ represents the decay rate of the eigenmatrix $\hat{\rho}_{1}^{(\nu)}$, Eq. (\ref{sym_breaking}) indicates that the state $\hat{\rho}_{1}^{(\nu)}$ never does decay. Here, let us also recall that the steady state ${\rho}_0$ has the $U(1)$ symmetry, whereas the state $\hat{\rho}_{1}^{(\nu)}$ has only $Z_{|\nu|}$ symmetry. Therefore, $-{\rm Re}[\Delta_{\nu(\neq0),1}]\rightarrow0$ means that once $U(1)$ symmetry is broken and a $Z_{|\nu|}$ symmetric state is realized by an infinitesimal symmetry breaking perturbation, the $U(1)$ symmetry is not recovered with a finite time. In this sense, $-{\rm Re}[\Delta_{\nu(\neq0),1}]\rightarrow0$ in the thermodynamic limit $\beta\rightarrow0$ could be interpreted as the ``symmetry breaking" in the ordered phase ($P>P_{\rm th}$).

\section{Discussion}
First, we discuss the possibility and difficulty of the experimental realization of high-$\beta$ class-A lasers. As Fig. \ref{fig:beta}(b) indicates, $g^{(2)}(0)$ is unity for any pump power in a class-A laser with $\beta=1$. Practically, this regime is very attractive, for example, for optical communication because Poissonian light can be emitted with an arbitrary low pump power. Unfortunately, for the following reasons, the experimental realization of this regime is not easy.
Both class-A and high-$\beta$ conditions can be experimentally realized by an ultra- high Q cavity and Purcell enhanced carrier emission into a cavity mode in state-of-art semiconductor microcavities. The major obstacle may lie in the assumption of the zero lower level population in emitters. The effect of the finite lower level population is effectively introduced in the rate equations  (\ref{eq:REp}) and (\ref{eq:REn}) by replacing the stimulated emission term $\beta\gamma_\|nN$ with $\beta\gamma_\|n(N-N_0)$ \cite{Rice1994}, where $N_0$ is called a ``carrier transparency number". We found that, for lasers with $\beta\simeq1$, even a small number of $N_0$ (for example $N_0=10$) breaks the class-A condition.
Even though the condition $N_0\simeq0$ is possible in solid state lasers or atomic lasers as considered in this paper, this condition is technically almost impossible in semiconductor lasers. Therefore, the thresholdless laser ($\beta=1$) in the class-A limit shown in Fig. \ref{fig:beta} should be regarded as a thought experiment, which clarifies the physical meaning of $\beta$. 

Second, we briefly discuss the laser phase-transition analogy in class-B lasers. As we explained in Section 2A, in class-B lasers, the adiabatic elimination of the atom (carrier) degree of freedom cannot be possible, and  consequently equilibrium description is impossible (the detailed balance condition is violated). However, there are still various similarities between class-A and class-B lasers. For example, even in class-B lasers, the transition of $g^{(2)}(\tau)$ from 2 to 1 and linewidth narrowing can be observed. Of course, both class-A and class-B lasers with the same $\beta$ exhibit identical pump-input and light-output curves. Therefore, qualitatively the same figure as Fig. \ref{fig:beta} can be obtained for class-B lasers. However, photon statistics for class-B lasers depend on both $\beta$ and ratio $\gamma_\|/\gamma_c$, which is in contrast to class-A lasers whose photon statistics are characterized solely by $\beta$. Importantly, class-B lasers with $\gamma_\|/\gamma_c\ll1$ are known to emit super-Poissonian light even when their pump powers are far above the lasing thresholds \cite{Druten2000,Takemura2019,Wang2020}. Now, following questions arise. Does the laser-phase transition analogy hold for class-B lasers? What is the role of $\beta$ in class-B lasers? These questions are beyond the scope of this paper. However, we speculate that, to answer these questions, the Liouvillian spectrum analysis will be a powerful tool because it is applicable even for a quantum system without the detailed balance condition. It is of particular interest where or not the laser linewidth corresponds to the Liouvillian gap even for class-B lasers. Furthermore, the behavior of the Liouvillian gap in the limit $\beta\rightarrow0$ is important in terms of the quantum phase transition violating the detailed balance condition.

\section{Conclusion}
We investigated photon statistics, linewidth, and the laser-phase transition analogy for low- and high-$\beta$ lasers in the class-A limit. While the spontaneous emission coupling coefficient $\beta$ is commonly used in class-B lasers to represent the fraction of photons spontaneously emitted into a lasing mode, it is rarely discussed in relation to class-A lasers, where only the photon degree of freedom is important because the photon lifetime is much longer than the other lifetimes. Our main results can be summarized as following.

We demonstrated that the spontaneous emission coupling coefficient $\beta$ can be defined even in the class-A limit and that $\beta$ uniquely characterizes both static and dynamic photon statistics. Interestingly, class-A lasers with $\beta=1$ do not exhibit any phase-transition-like behavior in photon statistics. To study the laser linewidth of low- and high-$\beta$ class-A lasers, we introduced $\beta$ in the well-known Scully-Lamb full master equation. We found that even class-A lasers with $\beta=1$ display  linewidth narrowing with increases in pump power. Finally, we discussed the laser-phase transition analogy with the Liouvillian structure of the Scully-Lamb master equation. We demonstrated that the Liouvillian gap of the Scully-Lamb master equation is equivalent to its laser linewidth, which can be interpreted in terms of a second-order quantum dissipative phase transition with continuous phase symmetry [the $U(1)$ gauge symmetry].

\section{Acknowledgments}
We thank V. Bastidas and K. Takata for fruitful discussions.

%

\end{document}